\documentclass[aps,pra,a4paper,showpacs,twocolumn]{revtex4}
\usepackage{amssymb}

\usepackage{amsmath}
\usepackage{bm}
\usepackage{graphicx}
\usepackage{ulem}

\begin{document}
\title{Optical gratings induced by field-free alignment of molecules}
\date{\today}

\author{A.Rouz\'ee, V.Renard, S.Gu\'erin, O.Faucher, and
B.Lavorel}
\affiliation{Laboratoire de Physique de l'Universit\'e de
Bourgogne, UMR CNRS 5027, BP 47870, 21078 Dijon Cedex, France}
\begin{abstract}
We analyze the alignment of molecules generated by a pair of crossed ultra-short pump pulses of different polarizations by a technique based on the induced time-dependent gratings. Parallel polarizations yield an intensity grating, while perpendicular polarizations induce a polarization grating. We show that both configurations can be interpreted at moderate intensity as an alignment induced by a single polarized pump pulse. The advantage of the perpendicular polarizations is to give a signal of alignment that is free from the plasma contribution. Experiments on femtosecond transient gratings with aligned molecules were performed in CO$_2$ at room temperature in a
static cell and at 30 K in a molecular expansion jet.
\end{abstract}

\maketitle

\section{Introduction}

Excitation with non-resonant high power laser pulses has become a
powerful technique for preparing aligned molecules \cite{Stap}. With ultra-short
laser pulses, typically of less than 100 fs pulse duration, molecules
periodically realign after the pump extinction, showing revivals
in the observed signal. Various methods are employed to measure
such post-pulse molecular alignment. They make use either of
ionization-fragmentation combined with an imaging technique
\cite{Stap,Sakai} or of non-linear optical properties
\cite{Renard2003,Renard2005}. In the methods of the latter category,
the depolarization or the spatial deformation of a probe pulse is
measured as a function of time elapsed after the aligning pulse.
The effect is based on  the time-dependent non-linear
contribution of alignment to the refractive index
\cite{Renard2003,Renard2005}. Transient grating experiments have
been also performed to explore alignment of molecules
\cite{Comstock,Stavros}. In these experiments, a transient grating
was produced by two synchronized pump pulses, and the amount of
light diffracted from a third probe pulse was
recorded at different delays. It appeared that the recorded signal exhibited the same
time behavior than other alignment signals, i.e. confirming the detection of
post-pulse molecular alignment. As in Refs.
\cite{Renard2003,Renard2005} a deformation of the measured signal
with increasing intensity has been noted. Whereas in Ref.
\cite{Comstock}, the change in the shape of the revivals has been
attributed to molecular changes and the apparition of a constant background
due to ionization and plasma formation, a more likely explanation for
this modification of DFWM transient shapes has been suggested
\cite{Lavorel}: the background is due to permanent alignment and
the alteration of revivals is a consequence of heterodyning with
this background. In fact, this had been already pointed out for
all experiments with homodyne detection \cite{Renard2003}. Stavros
et al. \cite{Stavros} corroborated this explanation in the case of
DFWM for O$_2$ molecules. At the highest intensity investigated in
Ref. \cite{Stavros}, the possible contribution of an ion grating
has also been suggested.

The aim of the present work is to investigate the mechanisms involved in high intensity DFWM experiments by using two
types of gratings (intensity or polarization grating) produced by two
crossed beams.
Furthermore, it will be shown that quantitative measurement of alignment with both configurations 
is possible with this technique.

\section{Experimental setup}

The experimental setup uses the well known femtosecond
degenerate four-wave mixing technique (DFWM) also called femtosecond transient
grating spectroscopy  \cite{Brown}. It consists of
exciting the molecular sample by two synchronized pump pulses at
800 nm (pulse duration of 90 fs) focused and crossed at a given
angle $\Theta$ (Fig.1). They are both derived from a chirped pulse amplified
Ti:Sapphire system working at 20 Hz. The energy and the polarization
direction of the two pump pulses are controlled by means of a half-wave plate and a polarizer. A third probe pulse, time delayed with
respect to the pumps, is mixed with the others. The probe beam sees a transient grating and is
diffracted. The amount of diffracted light is monitored as a
function of the time delay with a photomultiplier. The sample is
a CO$_2$ gas, either at room temperature in a static cell, or in a molecular supersonic jet for lower rotational temperatures. Both
beams are focused and collimated with $f=300$ mm lenses. It is
noticed that, like in other all-optical methods, DFWM gives the
opportunity to work with a good sensitivity in a wide pressure range (typically from 1 mbar to a few bars). The role of the two pump pulses
is to excite the rotational coherence of the molecules. Two
different experiments have been performed which correspond to
parallel and perpendicular polarizations of the pump pulses. The
two cases are treated separately in the next sections.

\section{Theory}

\subsection{Parallel polarizations: linear total  polarization,
intensity grating}
\label{para polar}
\begin{figure}[tbph]
\centerline{\includegraphics[scale=0.4]{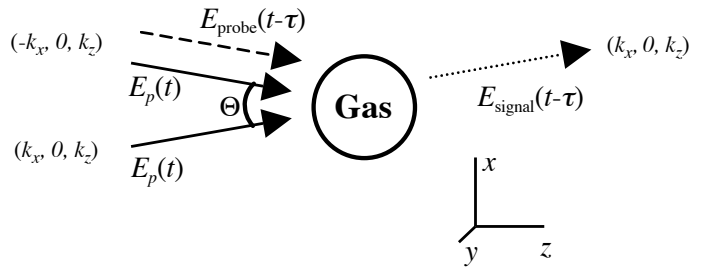}}
\caption{Experimental scheme for transient grating experiments
(see text). $E_p$ and $E_{\text{probe}}$ are respectively the pump and probe field envelope. The diffracted signal field envelope is $E_{\text{signal}}$ . } \label{scheme}
\end{figure}

We consider a simplified model for the formation of the grating. We assume
two plane waves, with wavelength $\lambda$,  propagating in the $(x,z)$ plane with wave vectors
$(-k_x,0,k_z)$ and $(k_x,0,k_z)$ (see Fig. 1). The small
cross-angle  between the two  pump beams is given by
$\sin(\Theta/2)=k_x\lambda/2\pi\approx\Theta/2$. With two pump pulses
polarized along the $y$-axis (of
vector unit $\vec{e}_y$), an intensity grating is formed with a
total electric field periodically modulated along the $x$-axis:
$\vec{E}_{\text{pump}}=2E_p(t)\cos(k_x x)\cos(\omega t
-k_z z)\vec{e}_y$, where $E_p(t)$ is the pump field
envelope (taken the same for both fields).  We are
facing therefore the case of molecules interacting with a non
resonant linearly polarized laser field, which has been extensively
studied (see Ref. \cite{Stap} and references therein). The
additional ingredient is the spatial modulation of the
total intensity. If we consider a weak probe pulse polarized parallel
to the pumps ($y$-axis) and which propagates with the wave vector
$(-k_x,0,k_z)$, the only component of the induced dipole at a position $\vec{r}=(x,y,z) $ whose expectation value does not average out is the $y$-component  : 
\begin{eqnarray}
\label{mu} \langle\vec{\mu}_{\text{ind}}\rangle_{\vec{r},t} & =
&\left[\bar\alpha +\Delta\alpha\left(
\langle\cos^2\theta_y\rangle_{\vec{r},t}-1/3\right)\right]\notag\\
&& \times E_{\text{probe}}(t)\cos(\omega
t+k_xx-k_zz)\vec{e}_y,\quad
\end{eqnarray}
where $\bar\alpha$ is the mean polarizability,
$\Delta\alpha=\alpha_{\parallel}-\alpha_{\perp}$ is the difference
of polarizability parallel and perpendicular to the molecular
axis, $E_{\text{probe}}(t)$ is the probe envelope, and
$\theta_y$ is the angle between the molecular axis and the common
polarization direction ($y$-axis). The expectation value
$\langle\cos^2\theta_y\rangle_{\vec{r},t}$ is a measure of alignment
and is a function of the total pump intensity at the position $\vec{r}$. Consequently, it has the same spatial period as the total intensity. It
also depends on the time delay between the pump and probe pulses.
The signal electric field measured by the
photo-detector is then obtained by summation over the interaction
volume of spherical waves radiated by the dipole. The first term of the right hand
side of Eq. (\ref{mu}) describes the effect of the linear index of
refraction on the probe beam and is not interesting for our
purpose.  The second term is periodic and leads to constructive
interference along  symmetric phase-matching
directions  $x/z=(2n-1)k_x/k_z$, with $n$ the diffraction order. Only the order 1 is considered experimentally (see Fig.
\ref{scheme}). The experimental signal measured at a position $\vec{r_0}$, proportional to the modulus squared of the total electric field, is interpreted as the diffraction
of the probe beam by the transient  grating produced by
the pumps

\begin{equation}
\label{signal}
I_{\text{sig}}(\vec{r_0},t)\propto |\vec{E}_\text{sig}(\vec{r_0},t)|^2 \propto |\int\int\int_V dr{\cal
R}[\vec{E}_{\text{rad}}(\vec{R},t)]|^2,
\end{equation}
with $ \vec{E}_{\text{rad}}$ the spherical
electric field radiated by the dipole moment 
\begin{equation}
\label{erad}
\vec{E}_{\text{rad}}(\vec{R},t)=\frac{k^2}{4\pi\epsilon_0}
\langle\vec{\mu}_
{\text{ind}}\rangle_{\vec{r},t}  \ \frac{e^{ikR}}{R}
\end{equation}
and with  $\vec{R}=\vec{r_0}-\vec{r}$,
$x/x_0,y/y_0,z/z_0\ll 1$, $k \simeq \omega/c$.

At low and moderate intensities (i.e. below the saturation of
alignment), numerical simulations show that $
\langle\cos^2\theta_y\rangle_{\vec{r},t}-1/3$ is approximately
proportional to the total pump intensity $4I_0\cos^2(k_x
x)=2I_0(1+\cos2k_x x)$, where $I_0$ is the peak intensity
of a single pump (see Appendix \ref{linear}). The constant term 2$I_0$ corresponds to the zeroth order of diffraction and is responsible for nonlinear variation of the refractive index seen  by the probe pulse. The second term 2$I_0\cos2k_xx$ leads to diffraction orders $n=\pm1$. It is shown in Appendix \ref{linear} that :
\begin{equation}
\langle\cos^2\theta_y\rangle_{\vec{r},t}-1/3\approx
\left(\langle\cos^2\theta_y\rangle_t-1/3\right)(1+\cos2k_xx),
\end{equation}
with
\begin{equation}
\label{sumcosappro} \langle\cos^2\theta_y\rangle_t-1/3\approx
2\xi_0[\delta+\kappa f(t)],
\end{equation}
and
\begin{equation}
\xi_0 =\Delta\alpha/4\hbar\int dt E_p^2(t),
\end{equation}
the intensity area of a single pump. $ \langle\cos^2\theta_y\rangle_t$ corresponds to the alignment that would be induced by a single linearly polarized pulse of intensity $2I_0$. For a given molecule and
temperature, $\delta$ and $\kappa$ are constant ($\delta \text { and } \kappa \to 0$ at
low intensity), and $f(t)$ is a specific function
independent of $\xi_0$. The summation over the volume  (Eq.(\ref{signal})) gives a signal of intensity :
\begin{equation}
\label{siga}
I_\text{sig}(t)\propto\left(\langle\cos^2\theta_y\rangle_t-1/3\right)^2,
\end{equation}
where the spatial dependence has now disappeared. The
measured signal (order 1) can thus be calculated for any pump-probe time delay,
solving the time-dependent Schr\"odinger equation
\cite{Renard2003} using a single linearly polarized pump beam of
intensity $2I_0$. This takes into account the fact that the spatial distribution of the total intensity is modulated between 0 and $4I_0$.

In other words, the sample of aligned molecules presents a
nonlinear index of refraction, which depends on the
pump intensity. As the intensity along the grating is modulated by optical
interferences, a spatial modulation of refractive index is formed.
This gives a refractive index grating and a signal directly related to the post-pulse
alignment, as shown by Eq. (\ref{siga}).

\subsection{Perpendicular polarizations: elliptic total
polarization, polarization grating}

For perpendicular polarizations, with one pump aligned at $-45^{\circ}$ and the
other at $+45^{\circ}$ with respect to the $y$-axis, the resulting
field is in general elliptically polarized. The state of
polarization of the total pump electric field depends on the spatial
position and alternates between linear and circular, forming a polarization grating.  In
a grating period, the total electric field varies from linear (along the
$y$-axis), circular (right), linear (along the $x$-axis), to circular
(left), with intermediate elliptic polarizations. Furthermore, its
magnitude varies periodically along the $x$-coordinate and is
maximum for the linear polarization and minimum for the circular
one. Since $\Theta$ is a small angle, the total pump field is
indeed polarized in the $(x,y)$ plane as
\begin{eqnarray}
\vec{E}_{\text{pump}}&=&\sqrt{2}E_p(t)[A(x)\sin(\omega t-k_z
z)\vec{e}_x\notag\\
&&+B(x)\cos(\omega t -k_z z)\vec{e}_y]
\end{eqnarray}
with $A(x)=\pm\sin(k_xx)$, $B(x)=\cos(k_xx)$, and $E_p(t)$ the pump
envelope. The interaction Hamiltonian reads \cite{Daems}:
\begin{subequations}
\label{ell}
\begin{eqnarray}
H_\text{int}&=&-\frac{1}{2}E^2_p(t)\Delta\alpha\sin^2\theta_z\notag\\
&&\times\left[\left(A^2(x)-B^2(x)\right)\cos^2\phi_z+B^2(x)\right]\\
&=&-\frac{1}{2}E^2_p(t)\Delta\alpha\notag\\
&&\times\left[A^2(x)\cos^2\theta_x+B^2(x)\cos^2\theta_y\right],\qquad
\end{eqnarray}
\end{subequations} where now $\theta_i$ ($i=x,y,z$) is the angle between the
molecular axis and the $i$-direction, and $\phi_z$
the corresponding azimuthal angle with respect to the
$z$-direction. We have used the identity
$\cos^2\theta_x=\sin^2\theta_z\cos^2\phi_z$.
 The quantities that do not
depend on $\theta_i$ nor $\phi_z$ have been omitted in
Eqs. (\ref{ell}) since they lead to irrelevant global
phases.

The wave function $\psi(x)$ can be estimated from the
simulation of the time-dependent Schr\"odinger equation including
the interaction (\ref{ell}), and with given functions
$A(x)$ and $B(x)$. The expectation value of the dipole induced by a $+45^{\circ}$
polarized probe can be then calculated at a given spatial position $r$. The induced
dipole whose component is filtered by a polarizer set at
$-45^{\circ}$ is given by:
\begin{eqnarray}
\label{mu3}
\langle{\mu}_{\text{ind}}\rangle_{\vec{r},t} & = & \frac{1}{2}
\Delta\alpha E_{\text{probe}}(t)\cos(\omega
t+k_xx-k_zz)\notag\\
&&\quad\times(\langle
\cos^2\theta_x\rangle_{\vec{r},t}-\langle\cos^2\theta_y\rangle_{\vec{r},t}).
\end{eqnarray}
It is shown in Appendix  (\ref{elliptic}) that the preceding equation can be approximated by :
\begin{eqnarray}
\label{mu2} \langle{\mu}_{\text{ind}}\rangle_{\vec{r},t} & \approx
& \Delta\alpha E_{\text{probe}}(t)\cos(\omega
t+k_xx-k_zz)\notag\\
&&\times\frac{3}{2}(A^2 (x)-B^2 (x))\xi_0(\delta+\kappa f_{\ell}(t)),
\end{eqnarray}
where we have considered that the Hamiltonian (\ref{ell}) can be
written as a superposition of two Hamiltonians corresponding
to linear pump polarizations along $x$- and $y$-directions, respectively.
The term $(A^2 (x)-B^2 (x))$ in (\ref{mu2}) can be rewritten as $\frac{1}{2}((1-\cos2k_xx)-(1+\cos2k_xx))$. It follows that the induced dipole can be considered as created by two out of phase intensity gratings with a spatial modulation $(1\pm\cos2k_xx)$. This has been initially demonstrated at low intensities when the perturbation theory
applies \cite{Fourkas} : a polarization grating can be decomposed in two out of phase intensity gratings with perpendicular polarizations. One result of the present work is to show that this simple decomposition can be extended beyond the perturbation regime at moderate intensities.
As in Section \ref{para polar}, the calculation of the observable
requires the summation of the radiated spherical electric fields
over the interaction volume. We consider the signal
produced along the phase-matching direction $x/z=k_x/k_z$ (order
1) and find that the associated signal can be obtained by
calculating numerically
$\langle\cos^2\theta\rangle_t-1/3$ using a linearly polarized
field at an intensity $I_0$ since the induced dipole moment (\ref{mu2}) is modulated between 0 (circular polarization, $A=B$) and a maximum value corresponding to a pump linear polarization with an intensity 2$I_0$.

\section{Results}

\subsection{Parallel polarizations}

For parallel polarizations, optical interferences
between the two pumps lead to an intensity grating. The
interaction Hamiltonian can thus be directly derived from earlier
studies of alignment with a linear polarization. The molecular
alignment reflects the total intensity and follows the fringe
pattern. An example is shown in Fig. 2 for pump intensities $I_0 = 19$ TW/cm$^2$. As in the polarization
technique, the modification of the experimental signal with respect to the weak field signal, is interpreted as alignment revivals
heterodyned by the permanent alignment.
The simulations (Eq. (\ref{siga})) should be performed at an
intensity taking into account the transverse profile of the two
pump beams (gaussian distribution) which was not considered in the theoretical section. This effect can be roughly accounted for by a factor 1/2 in the experimental
intensity, i.e. by comparing the theoretical intensity (about 20
TW/cm$^2$ for the best fit) to the peak intensity $I_0$ of only one pump beam (19
TW/cm$^2$). It is noted that, unlike in Ref.  \cite{Comstock}, here no molecular deformation is put forward to interpret the experimental signal.

\begin{figure}[tbph]
\centerline{\includegraphics[scale=0.26]{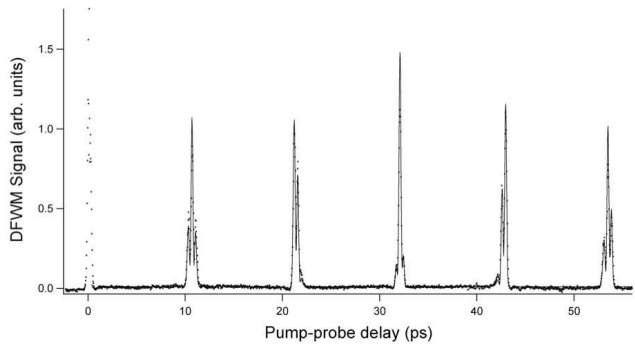}}
\caption{Recorded DFWM signal (dots) as a function of pump-probe
time delay in a static cell of CO$_2$ at room temperature for parallel
polarizations. Pressure is 0.1 bar and the peak intensity of
each pump is 19 TW/cm$^2$. Simulated signal [full line, Eq.
(\ref{siga})] is calculated at 20 TW/cm$^2$.} \label{Fig2}
\end{figure}

At higher intensity, a strong background is observed which
heterodynes and distorts the alignment signal. This phenomenon is
attributed to the contribution of a refractive index grating due to a spatial
distribution of electron density. Indeed, the electronic density
produced by ionization follows the spatial intensity distribution
of the pump with a lifetime of several nanoseconds \cite{Siders}.
As a consequence, the refractive index seen by the probe  is modified and a
grating is formed. The effect becomes very strong as the peak
intensity of the bright fringes approaches the ionization
saturation intensity (200 TW/cm$^2$ for CO$_2$). Some applications of this effect, which
limits the range of intensity for the present study of alignment,
is discussed in Ref. \cite{Loriot}.

\subsection{Perpendicular polarizations}

To overcome the intensity limitation discussed above, the
polarizations of the two pump beams have been crossed. In that
case, no optical interferences occur, and the intensity is
constant over the interaction volume, except for the overall
Gaussian distribution. A polarization grating is formed as discussed in the previous sections. Nevertheless, the instantaneous total intensity for the linear polarization is twice that of the circular one. As the ionization process is a function of the instantaneous
intensity, the ionization rate and the electron density reflect the periodic
distribution of this quantity. A refractive index grating due
to the electrons is formed as in the linear case.  But as the ionization rate is the same for the two
linear (along the $x$ and $y$-axis) and for the two circular polarizations (left
and right), the grating has a periodicity twice that of the parallel polarizations case (A).
The angle of diffraction is doubled and the probe diffracted by this grating can be spatially resolved from the one diffracted by the alignment-induced grating. The experiment is therefore not sensitive to the
electron density grating.

\begin{figure}[tbph]
\centerline{\includegraphics[scale=0.35]{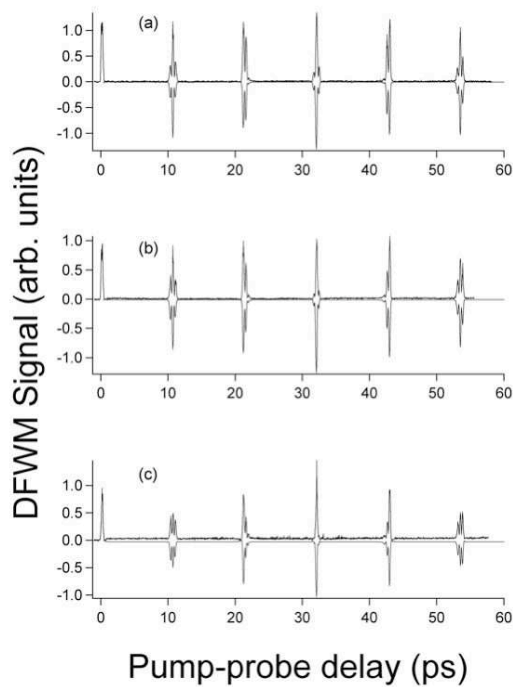}}
\caption{Experimental DFWM signal recorded as a function of the
pump-probe time delay in CO$_2$ at room temperature for
perpendicular polarizations. The pressure is 0.03 bar and pump
peak intensity is (a) 37 TW/cm$^2$, (b) 78 TW/cm$^2$, and (c) 135
TW/cm$^2$. The numerical simulations [Eq. (\ref{siga})] are shown
inverted for (a) 15 TW/cm$^2$, (b) 30 TW/cm$^2$, (c) and 55
TW/cm$^2$. These theoretical values compare well with $I_0$/2
(see text).} \label{Fig3}
\end{figure}

The Hamiltonian describing the interaction of the molecules with an
elliptic electric field as well as the induced dipole moment have
been given in the theoretical section. It has been shown that the
observed signal can be approximated by using the dipole moment (\ref{mu2}) at low and
moderate intensities. To simulate the experimental data, we thus
used the same routine as for parallel polarizations. Some
results are shown in Fig. 3. Again the transverse spatial profile of the pump pulses should be taken into account through a factor 1/2, i.e. the simulated intensity has to be compared to 1/2 times the maximum intensity of one pump beam. The degree of alignment,
quantified by $\langle\cos^2\theta\rangle_t$, is directly deduced
from simulations. Even though the higher experimental intensities correspond to saturation of $\langle\cos^2\theta\rangle_{r,t}-1/3$, the calculated intensity approaches the expected value (see Fig.\ref{Fig3}). The volume averaging of the signal tends to smooth out the saturation process of alignment. For much higher intensities,  discrepancies between calculated and experimental ($I_0/2$) intensity should appear, as shown in Ref.\cite{RenardPRA2004}. This effect is due to the strong saturation of  alignment at the center of the pumped volume. In that case, Eq.(\ref{mu2}) can not be used and the
observed signal is no more accurately simulated by Eq.(\ref{cosx-cosy}). Nevertheless, it would be possible, even though it
would be time consuming, to reproduce the experimental data by numerical simulation of Eqs.(\ref{ell}, \ref{mu3}).

One of the goals of the present work was to align cold molecules
in order to increase the efficiency of the process. To obtain low
temperatures, we used a pulsed supersonic jet in which the pressure
is only a few mbar, depending on the distance from the nozzle.
Before attempting an experiment, some preliminary studies at low
static pressure and room temperature have been performed. The
sensitivity was high enough to record signals at a pressure around 6
mbar of CO$_2$, with an excellent signal to noise ratio. The
supersonic jet has allowed us to decrease the temperature down to 30 K
(Fig. 4). The alignment that can be achieved at this temperature
with an experimental intensity of 47 TW/cm$^2$ corresponds to
$\langle\cos^2\theta\rangle=0.65$. This value is comparable to the one
achieved at higher intensity in our previous
study using a polarization technique at 60 K. But in the present
case the experiment is free from heterodyne contribution due to
spurious birefringence from optics.

\begin{figure}[tbph]
\centerline{\includegraphics[scale=0.40]{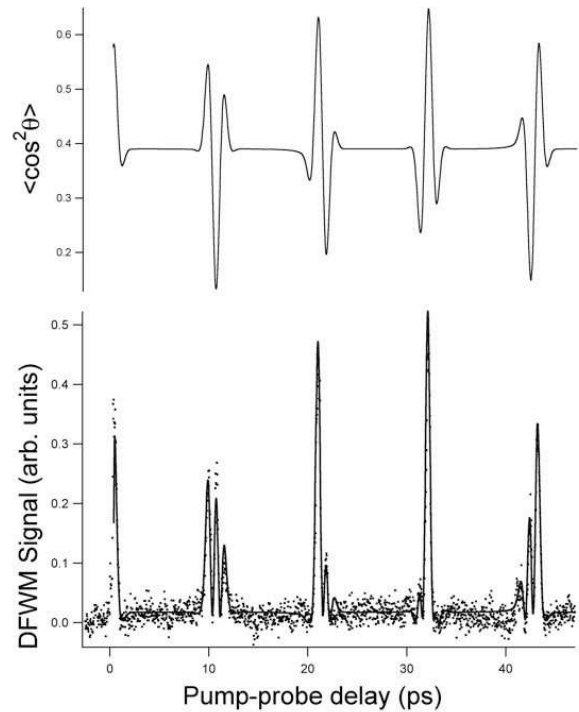}}
\caption{Recorded DFWM signal (lower panel, dots) as a function of
pump-probe time delay in CO$_2$ at 30 K in the supersonic jet for
perpendicular polarizations. Measured peak intensity of one pump
is 47 TW/cm$^2$. Simulated signal [lower panel, full line, Eq.
(\ref{siga})] is calculated at 25 TW/cm$^2$. The calculated time
dependence of $\langle\cos^2\theta\rangle$ is shown in the upper
panel.} \label{Fig4}
\end{figure}

The polarization grating (case B) can be discussed in terms of alignment
distribution, as for the intensity grating (case A). For example at a delay close to zero (modulo $T_R$),
the molecules are aligned along the linear polarized field ($x$ and
$y$-axis) and delocalized in the plane of polarization $xy$
(circular right and left). At time delay $T_R/4$ (modulo $T_R$), the molecules
are delocalized in the planes $xOz$ or $yOz$, or aligned along the
$z$-axis (circular right and left). Intermediate elliptic
polarization leads to intermediate state of alignment.

\section{Conclusion}

The main conclusions of this work are the following. DFWM experiments
with two synchronized spatially crossed femtosecond pulses allow to
produce transient gratings reflecting alignment of molecules. A contribution
due to ionization and plasma formation appears at high intensities
that heterodynes and therefore disturbs the experimental signal. With
perpendicular polarizations, the alignment signal can be isolated
from the plasma contribution. The experiments can thus be performed
at higher intensity keeping a good signal to noise ratio. DFWM experiments
prove to be very sensitive for probing low density and low temperature
samples such as those obtained with a supersonic jet. Furthermore, it
is shown that quantitative measurements of alignment are obtained
through simulations by using a simplified model, even in the case of perpendicular polarizations which was a priori non-trivial.

Using different polarization schemes would permit to create interesting
patterns of aligned molecules that could be used as a large frequency bandwidth spatial modulator.

\acknowledgments This work was supported by the Conseil R\'egional
de Bourgogne, the ACI \textit{photonique}, the CNRS, and a Marie Curie
European Reintegration Grant within the 6th European
Community RTD Framework Programme.
The authors thank H. R. Jauslin for helpful discussions.

\eject

\appendix

\section{Postpulse alignment for low and moderate intensities}
\label{Intermediate regime} We analyse in this appendix the
dependence of the postpulse alignment of linear molecules on the
pump field intensities below the intrinsic saturation regime. We
show in particular that (i) for a linearly polarized pulse the
baseline of $\langle\cos^2\theta\rangle$ obeys the following law:
quadratic and next linear, (ii) its shape is linear and (iii) the
alignment induced by an elliptically polarized field can be
analyzed in term of the decomposition into linearly polarized
fields.

\subsection{Linearly polarized field}
\label{linear}A linearly polarized field leads to a postpulse alignment
characterized by $\langle\cos^2\theta\rangle_t$, with $\theta$ the
angle between the molecular axis and the field polarization axis,
which can be generally characterized by
\begin{equation}
\label{sumcos}
\langle\cos^2\theta\rangle_t-1/3=C+\sum_J|a_J|\cos(\omega_Jt+\phi_J),
\end{equation}
where $C$ is a constant value corresponding to the permanent
alignment, and the second term, describing the rotational wave
packet revivals, is written in terms of Fourier components of the
amplitude $|a_J|$, the phase $\phi_J$, and the frequency $\omega_J$ (Raman
frequency of rotational transitions). In the present experimental conditions, the phases are roughly
constant: $\phi_J\approx-\pi/2$ \cite{RenardPRA2005}. Within the
sudden impulsive regime (i.e. for $\tau \ll h/B$ with $\tau$
the duration of the pulse and $B$ the rotational constant of the
molecule in Joule), where the pulse can be considered as a
$\delta$-function, the coefficients $C$ and $|a_J|$ depend on the
effective area
\begin{equation}
\xi =\Delta\alpha/4\hbar\int dt{E}^2,
\end{equation}
which is proportional to the peak field intensity $I$. For CO$_2$ molecules
interacting with a Gaussian pulse of a full-width at half maximum
$\tau_{\text{FWHM}}=0.1$ ps, we have $\xi\approx0.444\times
I[\text{TW/cm$^2$}]$.

\begin{figure}[h]
\centerline{\includegraphics[scale=0.7]{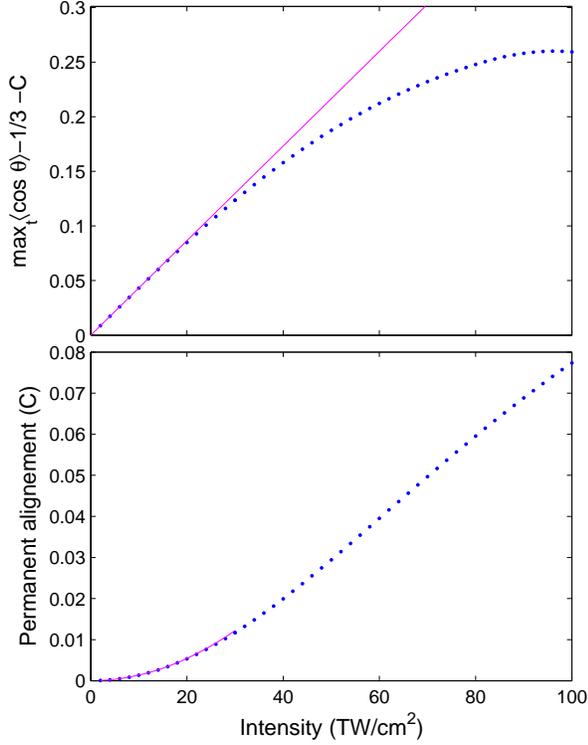}}
\caption{Permanent alignment $C$ and $\max_t
\langle\cos^2\theta\rangle_t-1/3-C$ as a function of the peak pump
intensity of $\tau_{\text{FWHM}}=0.1$ ps in CO$_2$ at $T=293$ K
(dotted lines). The former is quadratic (full fitting line)
approximately up to $I=30$ TW/cm$^2$ and linear for higher
intensities. The latter is linear (full fitting line) for moderate
intensities. Similar dependencies can be found for other linear
molecules and temperatures.} \label{C+max}
\end{figure}

Figure \ref{C+max} shows the dependence of the law (\ref{sumcos})
on the pump intensity (centrifugal distorsion has not been
considered in this study.) This allows us to define three regimes
of intensity: (i) low intensity associated to a quadratic
dependence of $C$ (here up to $I=30$ TW/cm$^2$), (ii) moderate
intensity associated to a linear dependence of $C$, and (iii) high
intensity associated to the saturation of $\max_t
\langle\cos^2\theta\rangle-1/3-C$. Below the saturation, one can
consider with a good approximation that $\max_t
\langle\cos^2\theta\rangle-1/3-C$ is approximately linear with
the field intensity. This leads for low intensities to
\begin{equation}
\label{sumcosappro1} \langle\cos^2\theta\rangle_t-1/3\approx \beta
\xi^2+\kappa \xi f(t)\approx \kappa \xi f(t),
\end{equation}
where, for a given molecule and temperature, $\kappa$ and $\beta$
are constant and $f(t)$ is a specific function independent of
$\xi$. We can notice that the regime of low intensities extends
the result of the perturbative regime, even if it is not itself a
perturbative regime (usually defined as a small population
transfer), since it can exhibit a non negligible alignment ($\max_t
\langle\cos^2\theta\rangle_t\approx 0.45$ for $I=30$ TW/cm$^2$).
We also have $\beta \xi^2\ll \kappa \xi$, which shows that the
permanent alignment is negligible for low intensities. For
moderate intensity, Fig. \ref{C+max} shows that the permanent
alignment is linear with respect to the peak field intensity, which leads to
\begin{equation}
\label{sumcosappro2} \langle\cos^2\theta\rangle_t-1/3\approx
[\delta+\kappa f(t)]\xi.
\end{equation}
We can thus conclude that at low and moderate intensities (i.e.
below the saturation of alignment),
$\langle\cos^2\theta\rangle_{t}-1/3$ is approximately proportional to $\xi$, i.e.
to the pump intensity $I$.

\subsection{Elliptically polarized field}
\label{elliptic}
We consider here the postpulse alignment induced by an
elliptically polarized field $\vec{ E}(t)={
E}(t)(A\cos(\omega t)\vec{e_x}+B\sin(\omega t)\vec{e_y})$,
$A^2+B^2=1$. The associated interaction reads
\begin{subequations}
\begin{eqnarray}
\label{int_ell} H_\text{int} &=&-\frac{\Delta\alpha}{4}{
E}^2(t)
\left[\left(A^2-B^2\right)\cos^2\theta_x-B^2\cos^2\theta_z\right]\qquad\\
&=&-\frac{\Delta\alpha}{4}{E}^2(t)
\left[A^2\cos^2\theta_x+B^2\cos^2\theta_y\right]
\end{eqnarray}
\end{subequations} with the particular cases of linear
polarizations along $x$ for $A=1$, along $y$ for $B=1$, and of
circular polarization in the $(x,y)$ plane for $A=B=1/\sqrt{2}$.

In the sudden impulsive regime, the propagator reads (the pulse
interacting with the molecule at time $t=0$)
\begin{equation}
U(t,0)=e^{-iJ^{2}Bt/\hbar} e^{i\xi A^2\cos^2\theta_x}e^{i\xi
B^2\cos^2\theta_y},
\end{equation}
where the irrelevant global phases have been omitted. This
propagator can be interpreted as the propagator associated to two
interactions with fields linearly polarized along $x$ of effective
area $\xi A^ 2$ and along $y$ of effective area $\xi B^2$. To calculate the expectation value (\ref{mu3}), we first consider ($\langle\cos^2\theta_i\rangle_t-1/3$) ($i=x,y,z$) as a superposition of two quantities corresponding to the action of the two fields :
\begin{eqnarray}
\label{sumobs}
\langle\cos^2\theta_i\rangle_t-1/3&=&(\langle\cos^2\theta_i\rangle_t-1/3)_{x-polar,A^2}\nonumber\\
&&+(\langle\cos^2\theta_i\rangle_t-1/3)_{y-polar,B^2},\nonumber\\
\end{eqnarray}
where the first term is calculated for the $x-$polarized field with a weighting factor $A^2$ and the second one for the $y-$polarized field with a weighting factor $B^2$. Beginning with $i=x$, the first term as estimated from (\ref{sumcosappro2}), is $\approx A^2 \xi(\delta+\kappa  f(t))$. For the $y-$polarized field, we would have by symmetry
$\langle\cos^2\theta_x\rangle_t=\langle\cos^2\theta_z\rangle_t$, and thus
$\langle\cos^2\theta_x\rangle_t=(1-\langle\cos^2\theta_y\rangle_t)/2$, using the property
$\cos^2\theta_x+\cos^2\theta_y+\cos^2\theta_z=1$. The second term is therefore $\approx -\frac{1} {2}B^2\xi(\delta+\kappa  f(t))$, using again \ref{sumcosappro2}. The summation in (\ref{sumobs}) gives $\langle\cos^2\theta_x\rangle_t-1/3 \approx \xi(A^2-B^2/2)(\delta+\kappa f(t))$.
Performing the same calculation for $i=y,z$, we finally find that in the intermediate field regime,
\begin{subequations}
\begin{eqnarray}
\langle\cos^2\theta_x\rangle_t-1/3 &\approx&
\xi(A^2-B^2/2)(\delta+\kappa f(t)),\qquad\\
\langle\cos^2\theta_y\rangle_t-1/3 &\approx& \xi(B^2-A^2/2)(\delta+\kappa f(t)),\qquad\\
\langle\cos^2\theta_z\rangle_t-1/3 &\approx& -\xi(\delta+\kappa
f(t))/2.\qquad
\end{eqnarray}
\end{subequations}
We have in particular that $\langle\cos^2\theta_x\rangle_t-1/3$ is
well approximated by the quantity $\xi(\delta+\kappa f(t))$
obtained from the interaction with a linearly polarized field of
peak intensity $I$ times the scale factor $(A^2-B^2/2)$. We
remark that $\langle\cos^2\theta_z\rangle_t-1/3$ does not depend
on the ellipticity. We also notice that
$\langle\cos^2\theta_y\rangle_t-1/3 \approx0$ when $B^2=A^2/2$,
i.e. $B^2=1/3,A^2=2/3$. This particular ellipticity allows the
directions $\theta_x$ and $\theta_z$ to play a symmetric role in
(\ref{int_ell}). This property has been used in Ref. \cite{Daems}
to demonstrate the optimal alternation of postpulse alignment.

For perpendicular polarizations, the observable $\langle
\cos^2\theta_x\rangle_{t}-\langle\cos^2\theta_y\rangle_{t}$ is
required. We obtain
\begin{equation}
\label{cosx-cosy}
\langle
\cos^2\theta_x\rangle_{t}-\langle\cos^2\theta_y\rangle_{t}\approx
\frac{3}{2}(A^2-B^2)\xi(\delta+\kappa f(t))
\end{equation}
We can conclude that a good approximation for
$\langle\cos^2\theta_x\rangle_{t}-\langle\cos^2\theta_y\rangle_{t}$
can be obtained in the intermediate regime by calculating
numerically $\langle\cos^2\theta_i\rangle_t-1/3$ with a single
field linearly polarized along $i$ of intensity $I$ and by
applying the scale factor $\frac{3}{2}(A^2-B^2)$.

\end{document}